\documentclass[%
superscriptaddress,
bibnotes,
amsmath,amssymb,
aps,
prl, 
noeprint ,
floatfix,
reprint
]{revtex4-2}
\usepackage{graphicx,xcolor}
\definecolor{darkblue}{RGB}{0,0,150}
\definecolor{nightblue}{RGB}{0,0,100}

\usepackage[
colorlinks,
citecolor=darkblue,
linkcolor=darkblue,
urlcolor=nightblue]{hyperref}

\usepackage[english]{babel}
\usepackage[babel,kerning=true,spacing=true]{microtype}

\usepackage{ulem}
\usepackage{physics,amscd,dsfont,wasysym,mathrsfs,mathtools}
\usepackage{multirow}
\usepackage{dcolumn}
\usepackage{comment}
\usepackage{xcolor}
\usepackage{bbm}

\usepackage{booktabs}

\begin{document}

\title{
Transverse voltage in anisotropic hydrodynamic conductors}

\author{Kaize Wang}\email{kaize.wang@mpsd.mpg.de}\affiliation{Max Planck Institute for the Structure and Dynamics of Matter, Hamburg 22761, Germany
}
\author{Chunyu Guo}\affiliation{Max Planck Institute for the Structure and Dynamics of Matter, Hamburg 22761, Germany
}

\author{Philip J.~W.~Moll}\email{philip.moll@mpsd.mpg.de}\affiliation{Max Planck Institute for the Structure and Dynamics of Matter, Hamburg 22761, Germany
}
\author{Tobias Holder}\email{tobiasholder@tauex.tau.ac.il}\affiliation{School of Physics and Astronomy, Tel Aviv University, Tel Aviv 69978, Israel
}
\date{\today}

\begin{abstract}
Weak momentum dissipation in ultra-clean metals gives rise to novel non-Ohmic current flow, including ballistic and hydrodynamic regimes. 
Recently, hydrodynamic flow has attracted intense interest because it presents a valuable window into the electronic correlations and the longest lived collective modes of quantum materials.
However, diagnosing viscous flow is difficult as the macroscopic observables of ballistic and hydrodynamic transport such as the average current distribution can be deceptively similar, even if their respective microscopics deviate notably. 
Based on kinetic Boltzmann theory, here we propose to address this issue via the transverse channel voltage at zero magnetic field, which can efficiently detect hydrodynamic flow in a number of materials. 
To this end, we show that the transverse voltage is sensitive to the interplay between anisotropic fermiology and boundary scattering, resulting in a non-trivial behavior in narrow channels along crystalline low-symmetry directions. 
We discuss several materials where the channel-size dependent stress of the quantum fluid leads to a characteristic sign change of the transverse voltage as a new hallmark of the cross-over from the ballistic to the hydrodynamic regime. 
\end{abstract}

\maketitle

\textit{Introduction.---} 
Hydrodynamic electron flow is a special transport regime which onsets when a rapid electron-electron scattering rate exceeds all other relaxation mechanisms. In recent years, its fundamental importance became apparent in understanding the longest lived collective modes of correlated electrons~\cite{Lucas2018b,Narozhny2022,Varnavides2023,fritz2024hydrodynamic}.
For example, viscous (hydrodynamic) correlations can shed light on electronic collective behavior, on the interacting phase diagram, and reveal unusual characteristics in the electron dynamics~\cite{Kovtun2005,Narozhny2016,Hartnoll2018}.
Hydrodynamic electron flow has also been discussed in connection to THz electromagnetic radiation in transistors~\cite{farrell2022terahertz,crabb2021hydrodynamic,dyakonov1993shallow}, 
ambipolar transports in semiconductors/semimetals~\cite{tan2022dissipation,nam2017electron,kukkonen1976electron} and fluid spintronics~\cite{matsuo2017theory,takahashi2016spin,takahashi2020giant,tatara2021hydrodynamic}.
The current conceptual frontiers of interacting Fermi liquids are found in understanding correlations on the nanoscale, either due to extreme scattering rates in strongly correlated electron systems such as high-$T_c$ superconductors, or due to nanoconfinement in heterostructures such as the nanoscopic channel sizes of current transistors. Advancing these goals necessitates experimental approaches sensing momentum diffusion in confined conductors.
The last few years saw tremendous experimental progress in the imaging~\cite{Ku2020,Vool2021,Ella2018,Kumar2022,AharonSteinberg2021,jenkins2022imaging} and characterization~\cite{Moll2016,Nandi2018,Gooth2018,fritz2024hydrodynamic} of the crossover regime between hydrodynamic and ballistic flow.  At the same time, these efforts have revealed that the flow profile of a quantum fluid in thin channels is rather ambiguous with respect to the dominant relaxation mechanism~\cite{Sulpizio2019}. 
While it is possible to alleviate this issue by choosing other geometries~\cite{AharonSteinberg2022,stern2022electron,Kumar2022,Bandurin2018} or measure at finite magnetic field~\cite{huang2021nonnegative,Scaffidi2017}, 
identifying more accessible characteristics sensitive to the ballistic-hydrodynamic crossover is highly sought after. The key point of this letter is to establish the transverse channel voltage in anisotropic conductors as such an accessible characteristic.

\begin{figure}
    \centering
	\includegraphics[width=1\linewidth]{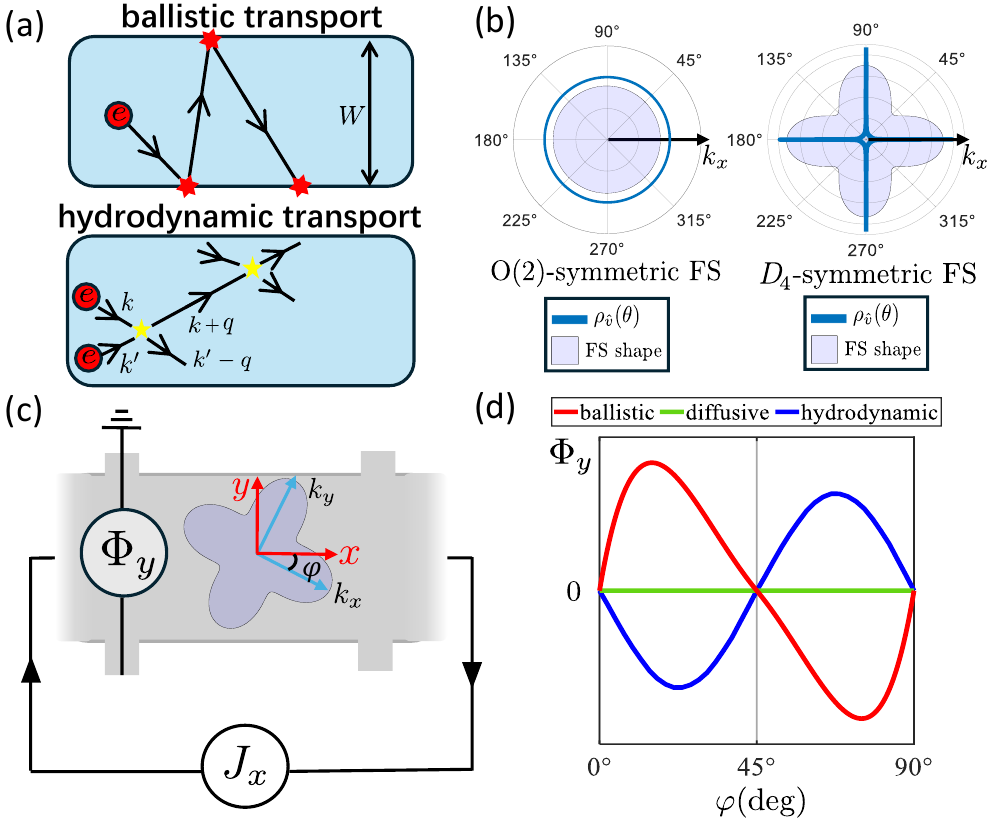}
	\caption{ (a) Ballistic and hydrodynamic transport regimes. 
    (b) Representative FS with continuous rotational symmetry and with anisotropy. $\rho_{\hat{v}}$ denotes the distribution of Fermi velocity directions, shown in a polar histogram.
  (c) A four point geometry for measurements of transverse voltages with misalignment angle $\varphi$ between the crystal principle axis and the channel direction. (d) Expected transverse voltage as a function of misalignment angle for a FS with $D_4$ symmetry.}
        \label{fig0}
\end{figure}

Electron transport in mesoscopic conductors is characterized by three characteristic length scales ~\cite{deJong1995}: The system size $W$, the momentum-relaxing (MR) mean-free path $\ell_{\mathrm{mr}}$, and the momentum-conserving (MC) mean-free path $\ell_{\mathrm{mc}}$. 
In very clean systems it holds that $W \ll \ell_{\mathrm{mr}}$ and Ohm's law no longer describes the local relationship between electric field and current. The system therefore becomes ballistic~\footnote{In this paper, the term ballistic transport
refers specifically to the bulk properties of the material. It does not imply the absence of boundary scattering, which is assumed
to be diffusive in our formalism.} if $W \ll \ell_{\mathrm{mr}} \ll \ell_{\mathrm{mc}} $  or hydrodynamic if $\ell_{\mathrm{mc}} \ll  W\ll \ell_{\mathrm{mr}}$ (Fig.~\ref{fig0}a). 
The physical tuning knob for $\ell_{\mathrm{mc}}$ is usually temperature, which controls the underlying dominant microscopic scattering mechanism~\cite{Gurzhi1968,Vool2021,Heilmann2024,Jaoui2018,Jaoui2021,Jaoui2022}. This mechanism can involve either non-umklapp electron-electron scattering or boson-mediated electron-electron interactions.

In previous studies, the crossover regime (Gurzhi regime) when all three length scales are comparable ($W\approx \ell_{\mathrm{mr}} \approx \ell_{\mathrm{mc}}$) has been studied theoretically by a Boltzmann kinetic approach with a Callaway two-rate ansatz for isotropic Fermi surfaces ~\cite{callaway1959model,deJong1995,Holder2019,Scaffidi2017}. 
The implicit assumption for this starting point is that the simplified isotropic description ought to capture the qualitative features of the ballistic-hydrodynamic crossover.  
However, this is far from obvious:
Experimentally, it is important to investigate the sensitivity of observables such as flow profiles against inevitable misalignments between the channel and the crystal axes. 
Since several candidate materials with hydrodynamic electronic transport are in fact metals with anisotropic Fermi surfaces~\cite{Moll2016,zaanen2019planckian,Vool2021}, it is critical to investigate the effect of symmetry breaking through the channel walls on 
hydrodynamic flow. Conversely, the Fermi surface (FS) shape can provide valuable insight for the ballistic-hydrodynamic crossover. Theoretical efforts to explore anisotropic effects in hydrodynamic conductors include broadband microwave spectroscopy~\cite{baker2023non,baker2024nonlocal}, a non-monotonic temperature and width dependence of the channel conductance~\cite{Cook2019}, and viscous flow profiles on a Corbino disk~\cite{Varnavides2020}. 

Here, we address this question by examining the transverse voltage in the absence of magnetic field. Our main focus is anisotropic systems (Fig.~\ref{fig0}b) lacking {{mirror symmetry}} due to a directional mismatch between the high symmetry axis of the FS and the geometric axis of the channel (Fig.~\ref{fig0}c). 
To describe these systems, we generalize the Callaway ansatz, finding a transverse voltage at zero magnetic field which is strongly sensitive to the microscopic scattering mechanism. 
Depending on the FS shape, a sign reversal of this transverse voltage can indicate that the system crosses over from ballistic to hydrodynamic flow (Fig.~\ref{fig0}d). 
We identify this strong dependence on the MC scattering rate as a competition of the transverse stress induced by the boundaries against the stress resulting from the bulk MR scattering.
{
Using the sign to identify the onset of hydrodynamic regimes offers several advantages. Firstly, it only requires a non-spatially resolved transport measurement of a single device. Secondly, the sign change of the transverse voltage is uncommon outside the hydrodynamic context, as it typically only relates to the carrier type of the material. However, we remark that the sign change does not mark a precise boundary point between the ballistic and hydrodynamic regimes.}

\textit{Collisional invariants for anisotropic Fermi surfaces.---}
We aim to describe the ballistic-hydrodynamic crossover using semiclassical kinetics. 
The starting point is the Boltzmann transport equation (BTE) for the distribution function $f$, given by
$(\partial_t + \mathbf{v}_{\mathbf{k}}\cdot \nabla_{\mathbf{r}} + \mathbf{F}\cdot \nabla_{\mathbf{k}}) f(\mathbf{r},\mathbf{p},t) = C[f(\mathbf{r},\mathbf{p},t)]$, where $C$ is the collision integral and $\mathbf{F}$ is the perturbation.  
The deviation from the equilibrium state is linearized by writing  $f(\mathbf{r},\mathbf{p},t) -f^0(\varepsilon_\mathbf{p})=  - E_F \partial_\varepsilon f^0(\epsilon) h(\mathbf{r},\mathbf{p},t)$, where $E_F$ is the Fermi energy and {$h$ is dimensionless non-equilibrium distribution function.} In the following, we consider the low temperature limit where $\partial_\varepsilon f^0(\varepsilon) \approx - \delta(\epsilon - E_F)$, suitable for materials with a large carrier density. 
For the sake of simplicity, we consider a single FS for which a bijection between angular variable $\theta$ and Fermi wave vector $\mathbf{p}_F$ exits. This allows to integrate out the radial momentum dependence of $\mathbf{p}$, leaving only an angular momentum variable $\theta$.  In more general scenarios, the FS arclength should be employed for each Fermi pocket separately, followed by a sum over multiple FSs.

\begin{figure*}
	\includegraphics[width=0.95\textwidth]{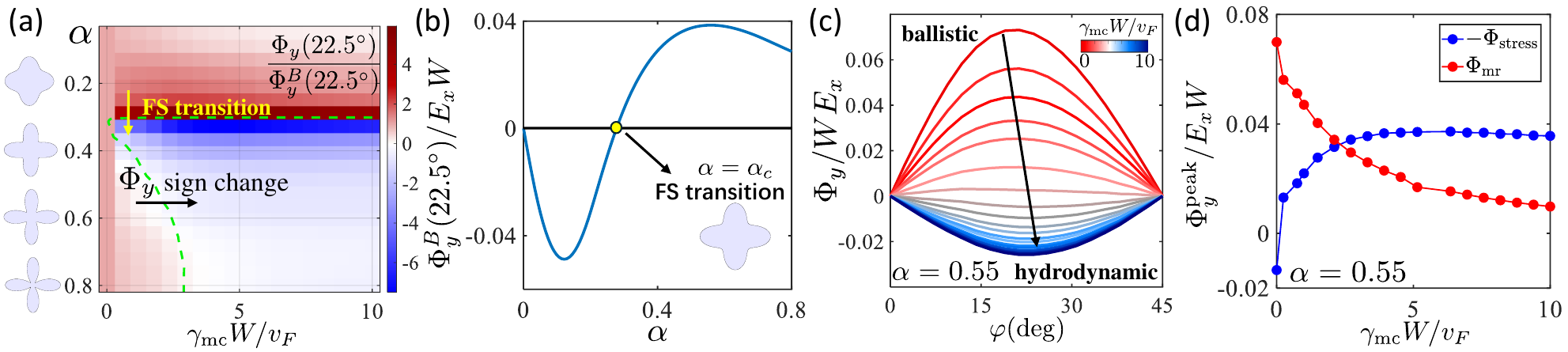}
	\caption{(a) Transverse voltage at $22.5^\circ$ normalized by the ballistic     
    value, shown as a function of both FS anisotropy $\alpha$ and dimensionless MC rate.
    Green dotted line: Interpolated boundary line when transverse voltage is vanishing. 
    Yellow arrow: FS transition at $\alpha \approx 0.3$, corresponding to a sign-change of ballistic transverse voltage, black  arrow: A line-cut for a fixed FS, where transverse voltage changes sign with increase MC rate. (b) Ballistic transverse voltage as a function of $\alpha$. Yellow dot: non-trivial zero of the ballistic transverse voltage signifying the FS transition. (c) Transverse voltage as a function of misalignment angle $\varphi$, parameters are taken as indicated by the black line cut in (a). From ballistic to hydrodynamic region, the transverse voltage flips its sign.   (d) Decomposition of the transverse voltage. As MC rate increases, MR contribution is decreasing to zero and the stress component is dominant. }
        \label{fig2}
\end{figure*}

We begin by defining a bra-ket notation for the BTE~\cite{Ledwith2019a,Cook2019,baker2023non}. Let $g(\mathbf{r},\theta,t)$ be a state function on the FS. Then the corresponding ket and inner product in two dimensions are defined by
$
    \ket{g(\mathbf{r},t)} = \int 
    {d\theta } \sqrt{A(\theta)} g(\mathbf{r} ,\theta,t) \ket{ \theta}
$
where $\bra{{\theta} }\ket{{\theta'} } = \delta(\theta- \theta')$  and the metric in phase space is $A(\theta) = { {p_F^2(\theta) E_F}/{{(2 \pi \hbar)^2|\mathbf{v}_F(\theta)\cdot \mathbf{p}_F(\theta)|}}}$. With this definition,  $\expval{g(\mathbf{r},t)} = \bra{h(\mathbf{r},t)}\ket{g(\mathbf{r},t)}$. We define the following modes with unit length: A particle mode called $\ket{c} \propto \int   d\theta \sqrt{A(\theta)} \ket{ \theta}$, corresponding to the particle number, and a momentum mode called $\ket{p_\mu} \propto \int   d\theta \sqrt{A(\theta)} p_\mu(\theta) \ket{ \theta} $, relevant for contractions which preserve momentum ~\cite{Ledwith2019a,Cook2019,baker2023non,Moll2016,deJong1995,supp}.   

To proceed, we take the Callaway two-rate ansatz~\cite{callaway1959model} 
for the collision integral $C$. The main idea is to restrict $C$ to a small set of longest-lived eigenmodes. 
For electron flow in a channel, these modes derive from particle number, which is conserved exactly; and momentum, which relaxes with a small scattering rate $\gamma_{\mathrm{mr}}$. We note that the assumption of isotropic scattering rates is approximate and most well-suited for weakly correlated materials which are good conductors~\cite{bachmann2022directional}.
Assuming that all other excitations relax at least as fast as these quasi-conserved quantities with rate $\gamma = \gamma_{\mathrm{mc}} + \gamma_{\mathrm{mr}}$~\cite{Cook2019,baker2023non}, one obtains two collisional integrals, 
\begin{eqnarray}
     C_{\mathrm{mr}} &=& -\gamma_{\mathrm{mr}}( \mathbbm{1} - \ket{c}\bra{c} ), \nonumber \\
    C_{\mathrm{mc}} &=& -\gamma_{\mathrm{mc}} \left(\mathbbm{1}  - \ket{c}\bra{c} - \sum_\mu \ket{p_\mu} \bra{p_\mu}\right).
\end{eqnarray}
As long as $W\ll v_F/\gamma_{\mathrm{mr}}$, this construction guarantees both particle number and (approximate) momentum conservation since $\bra{c}C_{\mathrm{mr/mc}}\ket{h} = 0$ and
$\bra{p_\mu}C_{\mathrm{mc}}\ket{h} = 0$.

Consider the current flow in a long narrow channel geometry (Fig.~\ref{fig0}) at zero magnetic field and zero temperature, which is translationally invariant in the  longitudinal direction (denoted $x$). Therefore, the distribution function only has a spatial dependence in $y$ direction. As a function of the two parameters $(y,\theta)$,
the BTE becomes, 
\begin{equation}
    D[h] = (C_{\mathrm{mr}} + C_{\mathrm{mc}}) \ket{h},
\end{equation}
where the drift terms are 
$
    D[h] =
    \partial_y \ket{v_y h} - E_x \ket{v_x} - E_y \ket{v_y}
$.
Here, $E_x,E_y$ are electric fields normalized by the Fermi energy, having the unit of a wave vector. The longitudinal electric field $E_x$ is fixed, while $E_y(y)$ is determined self-consistently. 
Rearranging terms, one obtains
\begin{align}
    C_0 \ket{h} &=\sum_\mu E_\mu \ket{v_\mu } 
    + \bigg [\gamma \ket{c}\! \bra{c} + \gamma_{\mathrm{mc}} \sum_\nu 
    \ket{p_\nu}\! \bra{p_\nu}\bigg ]\ket{h},
\end{align}
where $C_0 \ket{h} = \partial_y \ket{v_y h} + \gamma \ket{h}$. 
A diffusive boundary condition is imposed as
~\cite{deJong1995,Kiselev2019}
$
    h^\pm (\mp W/2,\theta) = 0
$
and 
$
    J_y(y) \equiv \bra{v_y}\ket{h} = 0.
$
\begin{figure*}[ht]
	\includegraphics[width=0.95\linewidth]{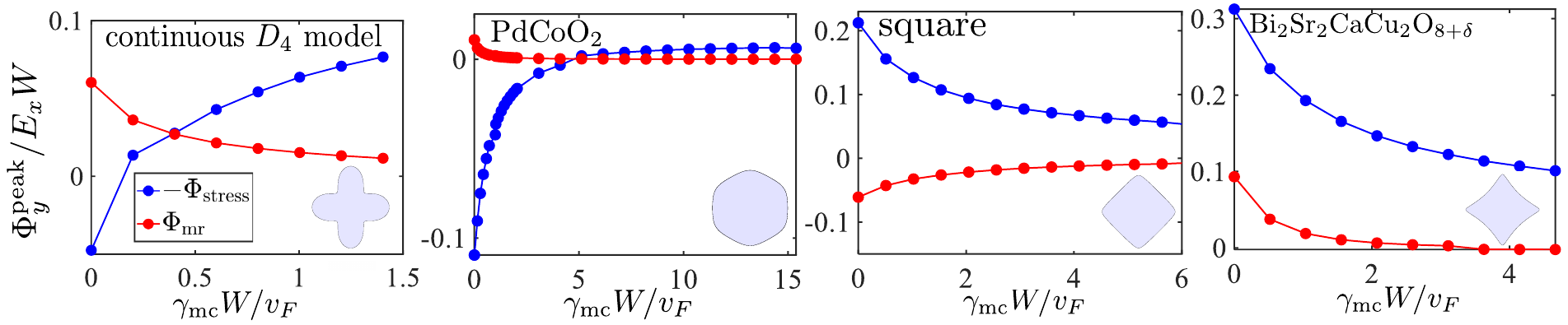}
	\caption{Decomposition of the transverse voltage of different models. Left to right: $D_4$ model with a  energy dispersion $\varepsilon(k)$ with $\gamma_{\mathrm{mr}}W/v_F = 0.1$~\cite{supp}, TB model for $\mathrm{PdCoO_2}$~\cite{takatsu2013extremely} with $\gamma_{\mathrm{mr}}W/v_F = 0.08$, TB square model with $\gamma_{\mathrm{mr}}W/v_F = 0.49$~\cite{supp},TB model for over-doped cuprate $\mathrm{Bi2212}$~\cite{markiewicz2005one}  with $\gamma_{\mathrm{mr}}W/v_F = 0.49$. Inset: Fermi surface shapes of different models.
    }
    \label{fig3}
\end{figure*}
Here,  $h^\pm(y,\theta)$ are $h(y,\theta)$ restricted to the  domain $\Theta^\pm = \{\theta|\mathrm{sign}[v_y(\theta)] = \pm 1\}$ respectively.  

In this form, we can solve the BTE self-consistently and obtain the transverse electric field as 
$
    E_y(y) = \bra{p_y}\ket{v_y}^{-1}(\bra{p_y}C_0
    \ket{h} - \gamma_{\mathrm{mc}} \sum_\mu 
    \bra{h}\ket{p_\mu}\bra{p_\mu}\ket{v_y})
$.
The transverse voltage is the integral $\Phi_y = \int dy \, E_y(y)$. 
We identify two different mechanisms in $\Phi_y$, originating from the stress due to ohmic dissipation and viscous dissipation, respectively. Denoting $\Phi_y = \Phi_y^{\mathrm{stress}} + \Phi_y^{\mathrm{mr}}$, it is
\begin{align}
    \Phi_y^{\mathrm{stress}} &=  \bra{p_y}\ket{v_y}^{-1} \bra{p_y}\ket{v_y h}\bigg|^{W/2}_{-W/2}\nonumber \\ 
    \Phi_y^{\mathrm{mr}} &=\bra{p_y}\ket{v_y}^{-1}\gamma_{\mathrm{mr}} \int_{-W/2}^{W/2} d y \bra{p_y} \ket{h}
    \label{eq:decom}
\end{align}

Here, $\Phi_y^{\mathrm{stress}}$ is related to the $yy$ component of the stress tensor $\Pi _{\mu \nu}\equiv \bra{p_\mu}\ket{v_\nu h}$ of the quantum fluid~\cite{Bradlyn2012,massignan2005viscous}, whereas $\Phi_y^{\mathrm{mr}}$ is proportional to the MR rate $\gamma_{\mathrm{mr}}$.

\textit{Ballistic-hydrodynamic crossover.---} 
In order to demonstrate the anisotropy effect of FSs in a tractable manner, we use the following parametrization for the Fermi wave vector $k_F$ for a $D_n$ FS
$k_F(\theta;\alpha,\varphi) = k_F^0 \left\{
    1 + \alpha \cos[n( \theta - \varphi)]\right\}$,
where $\theta$ is the angle between $\mathbf{k}_F$ and the crystal $k_x$-axis, $\varphi$ is the angle between the crystal and channel axes as shown in Fig.~\ref{fig2}a. $\alpha$ is the controlling parameter for the FS shape. With varying $\alpha$, the FS goes from circular to cross-shaped, and finally flower-shaped, as shown in Fig.\ref{fig2}a for $n=4$. 

Focusing on non-Ohmic region, we fix the MR rate $\gamma_{\mathrm{mr}}$ within the limit $\gamma_{\mathrm{mr}} W/v_F \lesssim 1$ and change the MC rate $\gamma_{\mathrm{mc}}$ to investigate the ballistic-hydrodynamic crossover. 
The transverse voltage at the misalignment angle $\varphi=22.5^\circ$ (i.~e. maximal misalignment) as a function of both FS shape and MC scattering rate is shown in Fig.~\ref{fig2}a. In order to track the sign change of the transverse voltage, we normalize by the ballistic voltage $\Phi_B$ with $\gamma_{\mathrm{mc}}=0$. 
For a fixed small $\alpha<\alpha_c$, the transverse voltage remains of the same sign irrespective of the value of $\gamma_{\mathrm{mc}}$. At the critical value $\alpha_c \approx 0.3$, the increasingly anisotropic FS leads to a sign change of the ballistic voltage (yellow arrow). 
Such a sign reversal is a direct consequence of the Fermi velocity distribution of different FSs, which is the only relevant factor in the ballistic limit.
In the supplemental material~\cite{supp}, we elucidate that this latter sign change is determined by the average $\expval{v_x \mathrm{sign}(v_y)}$. 

In contrast, for a fixed large $\alpha>\alpha_c$, with increasing MC scattering rate $\gamma_{\mathrm{mc}}$, the voltage exhibits an additional sign change between the ballistic region and the hydrodynamic region (cf. green dotted line in Fig. \ref{fig2}a). 
This latter sign change persists for all misalignment angles, as is shown 
in Fig. \ref{fig2}c. Thus, during the ballistic-hydrodynamic crossover, certain anisotropic FSs will cause the modulus of the transverse voltage to decrease, cross zero smoothly and change sign. Note that $\Phi_y(\varphi)$ can have a quite asymmetric shape, changing noticeably as a function of $\gamma_{\mathrm{mc}}$.
To investigate the origin of the transverse voltage,
$\Phi_y^{\mathrm{stress}}$ and $\Phi_y^{\mathrm{mr}}$ [Eq.~\eqref{eq:decom}] are shown in Fig.~\ref{fig2}d for the anisotropy parameter $\alpha=0.55$. $\Phi_y$ here is taken as the peak voltage in the whole range of $\varphi$ rotation at a fixed {$\gamma_{\mathrm{mr}}$ and varying with $\gamma_{\mathrm{mc}}$.  
We observe that in the hydrodynamic regime  ($\gamma_{\mathrm{mc}}W/v_F\gg 1$), the MR component is vanishing so that the  internal stress of the quantum fluid becomes the dominant contribution to the transverse voltage. Conversely, the dissipative stress dominates at small $\gamma_{\mathrm{mc}}$, therefore there exists a critical $\gamma^c_{\mathrm{mc}}$ where the two contributions cancel and the sign of transverse voltage $\Phi_y$ reverses.

\textit{Material examples.---} Here we present some instructive examples of anisotropic FSs which additionally contain anisotropic Fermi velocities. 
We find that both a continuum model and tight-binding (TB) examples can exhibit a behavior very similar to the one discussed for the simplified scenario, thus confirming the robustness of our diagnosis. Namely, we focus on the ultra clean material $\mathrm{PdCoO_2}$~\cite{Moll2016}, and the overdoped strongly correlated cuprate $\mathrm{Bi_2Sr_2CaCu_2O_{8+\delta}(Bi2212)}$~\cite{zaanen2019planckian}, as in both non-Ohmic transport has been observed. 
For comparison, we also include a square lattice with nearest-neighbor hopping.
The model details can be found in the supplemental material~\cite{supp},
the results of these calculations are shown in Fig.~\ref{fig3}. 
Not surprisingly, in all cases the MR component is diminishing when approaching the hydrodynamic limit. Thus in the hydrodynamic region, the only contribution to transverse voltage is the $yy$ component of the stress tensor $\Pi_{\mu\nu}$.
For the anisotropic $D_4$ model~\cite{supp}}, our calculation reveals a sign change of the transverse voltage with increasing MC scattering rate, serving as a direct indicator of hydrodynamic transport. The location of this crossover, however, depends on the FS shape, the channel width, and the el-el scattering rate. For the FSs of the square and Bi2212, for example, such a sign change would occur only at extreme scattering rates. Thus in a real material, these parameters may render the sign change inaccessible for realistic channel sizes. 

\begin{figure}
	\includegraphics[width=\columnwidth]{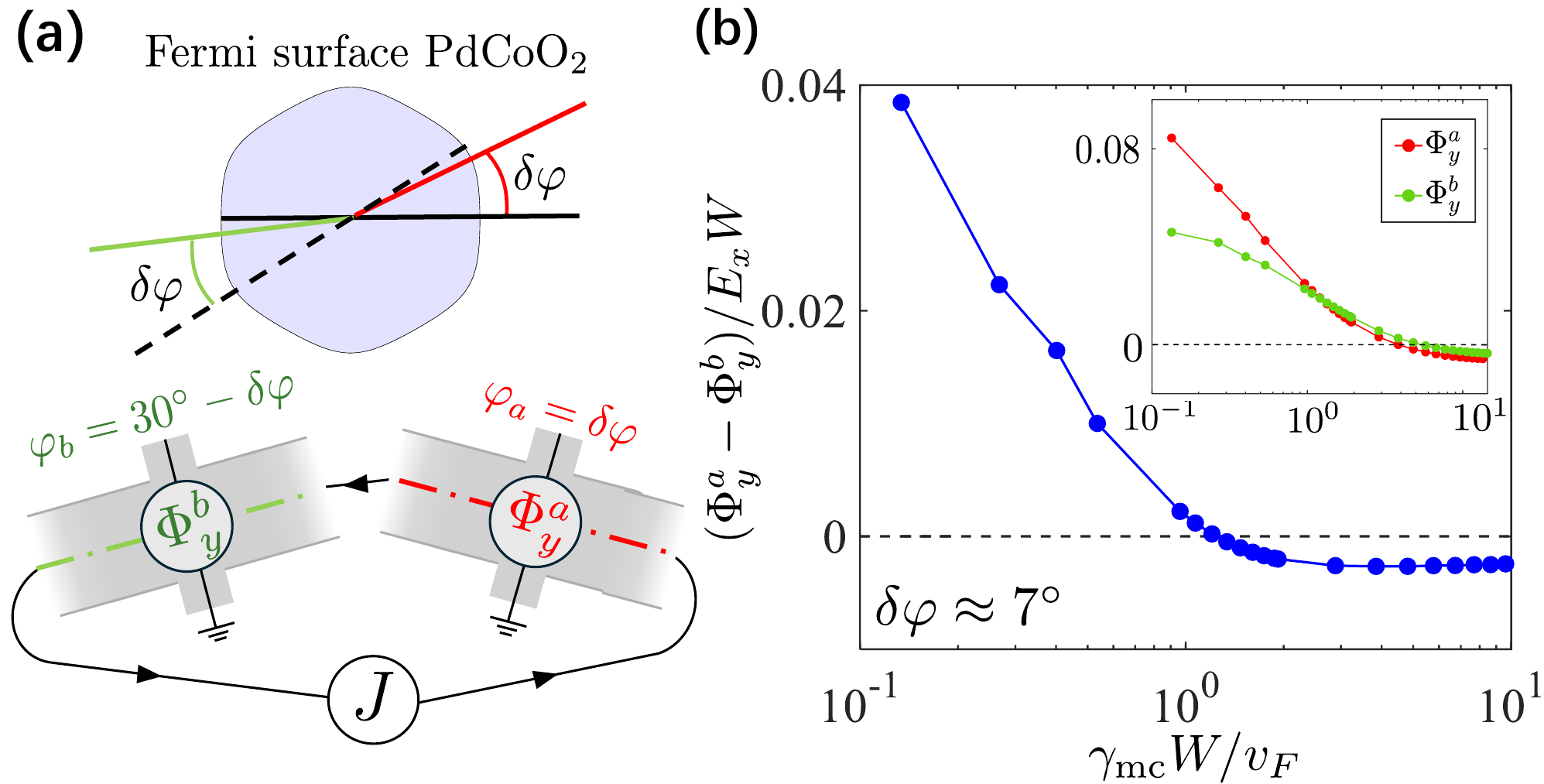}
	\caption{(a) Sketch for measurement of the difference of transverse voltage $\Phi_y^a - \Phi_y^b$ at misalignment angle $\varphi_{a} =  \delta \varphi$ and $\varphi_{b} = 30^\circ - \delta \varphi$ for $\mathrm{PdCoO_2}$. 
    (b) Difference of transverse voltage as function of MC scattering rate. Inset: $\Phi_y^a$ and $\Phi_y^b$ as a function of MC scattering rate.}
    \label{fig5}
\end{figure}

The diagnostic methodology can be extended to treat those situations, thus greatly expanding its range of applicability. The transverse voltage results from a competition between $\Phi_y^{\mathrm{stress}}$ and $\Phi_y^{\mathrm{mr}}$, which in general scale differently with the channel misalignment angle $\varphi$ (see asymmetry evolution in Fig.~\ref{fig2}c). Hence the transverse voltage difference between two channels symmetrically misaligned by an angle $\pm \delta \varphi$ to mutually adjacent mirror planes is a sensitive probe of emergent hydrodynamics (Fig~\ref{fig5}). Its asymmetry is actually most pronounced in the weakly hydrodynamic sector of small $\gamma_{\mathrm{mc}} W /v_F$ and a channel width dependence allows an estimation of $\gamma_{\mathrm{mc}}$ through a transport experiment. The required simple geometry of two canted bars is sketched in Fig.~\ref{fig5}a. Varying the channel width by a factor of 10 is straightforward lithographically, which will provide enough range to extract $\gamma_{mc}$ even when the sign change itself cannot be accessed.

\textit{Conclusion.---}
By employing a Callaway two-rate ansatz, we have identified a non-vanishing transverse voltage for a wide range of low-symmetry transport configurations. 
We find that in a $D_4$-symmetric system, a sufficiently anisotropic FS can lead to a sign change in the transverse voltage as the electron fluid crosses over from ballistic to hydrodynamic transport. 
We therefore propose that measuring the transverse voltage at zero magnetic field can be a viable way to distinguish different types of non-Ohmic transport. 
The prescribed phenomenology offers an alternative probe for the experimental investigation of unconventional charge transport beyond the analysis of the current flow pattern.
We reiterate that the measurement of transverse voltage seems particularly attractive because it does not require an external magnetic field or a local imaging of the current profile.
We believe that the sign change is observable with current devices as a function of either gate voltage or temperature using the geometry depicted in Fig.~\ref{fig5}. For example, in Ref.~\cite{Moll2016}, the estimated range of $\gamma_{\mathrm{mr}}/\gamma_{\mathrm{mc}}$ is $0.05\sim 0.2$, based on fits of the conductivity to an isotropic model.

As a limitation of our results, we note that the simplifications to the collision integral employed in this work cannot adequately capture the intermediate tomographic transport regime ~\cite{hofmann2023anomalously,kryhin2023collinear,hong2024superscreening,Ledwith2019a}.

The methodology developed here can be straightforwardly extended to investigate hydrodynamic crossovers in three-dimensional systems. Another promising direction for future work is the study of magneto-transport in anisotropic systems.

\begin{acknowledgments}
T.H.\ acknowledges financial support by the 
European Research Council (ERC) under grant QuantumCUSP
(Grant Agreement No. 101077020). 
\end{acknowledgments}
%


\end{document}


\title{Supplemental material to ``Transverse voltage in anisotropic hydrodynamic conductors"}

\author{Kaize Wang}\email{kaize.wang@mpsd.mpg.de}\affiliation{Max Planck Institute for the Structure and Dynamics of Matter, Hamburg 22761, Germany
}
\author{Chunyu Guo}\affiliation{Max Planck Institute for the Structure and Dynamics of Matter, Hamburg 22761, Germany
}
\author{Philip J.~W.~Moll}\email{philip.moll@mpsd.mpg.de}\affiliation{Max Planck Institute for the Structure and Dynamics of Matter, Hamburg 22761, Germany
}
\author{Tobias Holder}\email{tobiasholder@tauex.tau.ac.il}\affiliation{School of Physics and Astronomy, Tel Aviv University, Tel Aviv 69978, Israel
}
\maketitle

In the supplemental material (SM), we provide comprehensive details on the derivation of the Boltzmann kinetic approach for anisotropic Fermi surfaces. We introduce the a suitable Dirac notation for the phase space variables an account of the numerical solver. Then we derive the transverse voltage decomposition, discuss the ballistic limit and finally document the various model systems which we were investigated in the main text.

\date{\today}
    \appendix
    \tableofcontents
    \section{ Dirac notation in phase space and collision operators}
It is convenient to work with Dirac notations in Boltzmann transport equations to simplify things and avoid ambiguities~\cite{Cook2019,Ledwith2019a}. In the main text of the paper, we are considering two-dimensional Fermi liquids at zero temperature. Suppose the distribution function $f(\mathbf{r},\mathbf{p},t)$ is known. Then, for an arbitrary function $g(\mathbf{r},\mathbf{p},t)$ (here, $g$ can denote for example velocity, energy, displacement,...), the expectation value of $g$ is
\begin{equation}
      \expval{g} = \int \frac{d \mathbf{r} d \mathbf{p}}{(2\pi \hbar)^2} g(\mathbf{r},\mathbf{p},t)f(\mathbf{r},\mathbf{p},t). 
\end{equation}
We can define the density of the above quantity by 
\begin{equation}
     \expval{g(\mathbf{r})} = \int \frac{d \mathbf{p}}{(2\pi \hbar)^2} g(\mathbf{r},\mathbf{p},t)f(\mathbf{r},\mathbf{p},t).
\end{equation}
Within a linearized ansatz for the Boltzmann equation, the distribution function is usually parametrized as 
\begin{equation}
    f(\mathbf{r},\mathbf{p},t) = f^0(\epsilon(\mathbf{p})) - E_F \partial_\epsilon f^0 h(\mathbf{r},\mathbf{p},t),
    \label{eq:linear}
\end{equation}
where $h(\mathbf{r},\mathbf{p},t)$ is a continuous and differentiable function.
In a channel geometry, due to the translational invariance in the channel direction (direction $x$), the only remaining spatial variable is $y$. At zero temperaure, the Fermi function derivative is $\partial_\epsilon f^0 = -\delta(\epsilon_\mathbf{p} - E_F)$.
Putting these simplifications together, the density becomes ($\hbar=1$),
\begin{eqnarray}
	\langle g(y)\rangle = \int \frac{dp d \theta}{4\pi^2}  \,p f(y,\mathbf{p})g(y,\mathbf{p})
	&=&  \int \frac{dp d \theta}{4\pi^2} \, p f^{(0)}(\mathbf{p}) g(y,\mathbf{p})
	+ E_F \int  \frac{dp d \theta}{4\pi^2}  \,p \delta(\epsilon_\mathbf{p} - \mu) h(y, \theta)g(y,\mathbf{p}) \nonumber\\
	&=& E_F \int \frac{ d \theta}{4\pi^2}  \frac{p_F(\theta)}{|\partial_p \epsilon (\theta)|} h(y,\theta)g(y,\theta)\nonumber \\
	&=& E_F \int  \frac{d \theta}{4\pi^2}  \frac{p^2_F(\theta)}{|\mathbf{v}_F(\theta)\cdot \mathbf{p}_F(\theta)|} h(y,\theta)
	g(y,\theta)
 \label{eq:exp}
\end{eqnarray} 
where in the second line we assumed that $g$ is a non-equilibrium quantity which vanishes at equilibrium, while in the third line we use the fact that 
\begin{eqnarray}
    \partial_p \epsilon &=& \partial_{p_x} \epsilon \cdot \partial_p p_x + \partial_{p_y} \epsilon \cdot \partial_p p_y 
    =\frac{v_x p_x + v_y p_y}{p}.
\end{eqnarray}
From Eq.~\eqref{eq:exp}, we identify the measure in phase space as
\begin{eqnarray}
    A(\theta) = \frac{p_F^2 E_F}{4\pi^2|\mathbf{v}_F(\theta)\cdot \mathbf{p}_F(\theta)|}.
\end{eqnarray}
In this way we can define bra-kets, thereby writing Eq.~\eqref{eq:exp} as an inner product. 
Namely, for any function $g(y,\theta)$, the corresponding  ket and inner product are defined as 
\begin{eqnarray}
    \ket{g(y)} &\equiv& \int_0^{2\pi} d \theta \sqrt{A(\theta)} g(\theta,y) \ket{\theta} \nonumber \\
    \bra{\theta'}\ket{\theta} &=& \delta(\theta - \theta'). 
\end{eqnarray}
which leads to the very compact form $\bra{h}\ket{g}$ for Eq.~\eqref{eq:exp}. 
Likewise, for any linear operator $\hat{Q}$ one can write 
\begin{eqnarray}
    Q(y,y')
	&=& \int  d \theta d \theta' 
	{\sqrt{A(\theta )}}  Q(y,y',\theta,\theta') \frac{1}{\sqrt{A(\theta')}}  
	\ket{\theta}\bra{\theta'} 
\end{eqnarray}
Taking this convention, the distribution function and important modes like particle number $c$ momentum $p_i$ and velocity $v_i$ are given by 
\begin{eqnarray}
	\ket{h(y)} &=& \int d yd \theta\, \sqrt{A(\theta)}h(y,\theta)   \ket{\theta}, 
 \\\label{eq:modesc}
	\ket{c} &\equiv& \frac{1}{\sqrt{{N}}} 
	\int d \theta \sqrt{A(\theta)}\ket{\theta},
  \\\label{eq:modesp}
	\ket{p_i} &\equiv& \frac{1}{\sqrt{\expval{p^2_i}}} 
	\int d \theta\,\sqrt{A(\theta)} p_i(\theta)\ket{\theta}, 
 \\\label{eq:modesv}
	\ket{v_i} &\equiv& 
	\int d \theta \, \sqrt{A(\theta)}{v}_i(\theta)
	\ket{\theta}.
\end{eqnarray}
where $i=1,2$ denotes $x$ and $y$. Furthermore, 
\begin{eqnarray}
        p_i(\theta)&=& \mathbf{p}_F(\theta) \cdot \hat{x}_i, v_i(\theta)= \mathbf{v}_F(\theta) \cdot \hat{x}_i, \nonumber \\
	\expval{p^2_i} &\equiv& \int d \theta A(\theta)
	 p_i^2(\theta), \nonumber  \\
	N &\equiv&\int d \theta A(\theta)  .
\end{eqnarray}
The current therefore becomes
\begin{eqnarray}
	\expval{J_i} = \int dy\,\langle v_i\rangle(y) 
	\ket{y}=\bra{v_i}\ket{h}.
\end{eqnarray}

In order to construct the collision operator, it is necessary to first prove the orthogonality of the modes Eq.~\eqref{eq:modesp} and~\eqref{eq:modesv} for a Fermi surface possessing $C_{n\geq 4}$ symmetry. 
To this end, we will use the fact that $\varepsilon_\mathbf{k} = \varepsilon_{-\mathbf{k}} \Rightarrow p_F(\theta) = p_F(\theta \pm \pi), \mathbf{v}_F(\theta) = - \mathbf{v}_F(\theta\pm \pi),A(\theta ) = A(\theta\pm \pi )$.   
\begin{eqnarray}
	\bra{c}\ket{c} &=& 
	\frac{1}{N} \int_0^{2\pi} A(\theta )d \theta = 1,\nonumber \\
	\bra{c}\ket{p_i} &\sim& \int_0^{2 \pi} d \theta 
	\,A(\theta )p_\theta p_i(\theta) = \int_0^{\pi} d \theta \,A(\theta )p_\theta [p_{i}(\theta) + p_{i}(\theta -\pi )] = 0, \nonumber \\
	\bra{p_i}\ket{p_i} &=& 1.
\end{eqnarray}
These considerations would actually also apply for the less symmetric case  $C_2$ rotational symmetry. However, in order to prove the relation $\bra{p_i}\ket{p_j}=\delta_{ij}$, we do need  $C_{2m}(m\geq 2)$ symmetry, which allows use to employ that
\begin{eqnarray}
	p_F(\theta) = p_F\big(\theta + \frac{\pi}{m}\big)
	. 
\end{eqnarray}
The $C_{2m}$ symmetry for Fermi velocity therefore leads to
\begin{eqnarray}
	v_i({R \mathbf{k}}) 
	&=& \frac{\epsilon({R \mathbf{k}}+\delta k\hat{x}_i) - \epsilon(R \mathbf{k})}{\delta k}
	,\nonumber\\
	&=& \frac{\epsilon({R (\mathbf{k}}+\delta kR^{-1}\hat{x}_i)) - \epsilon(R \mathbf{k})}{\delta k},\nonumber \\
	&=& \frac{\epsilon(\mathbf{k}+[R^{-1}\hat{x}_i]\delta k) - \epsilon(\mathbf{k})}{\delta k}\nonumber\\
	&=& \mathbf{v}(\mathbf{k}) \cdot [R^{-1} \hat{x}_i]. 
\end{eqnarray} 
where $R$ is an arbitrary two-dimension rotation matrix. 
Generally for any $C_n(n\geq 3)$ symmetry, 
\begin{eqnarray}
	v_x\left(\theta + \frac{2 \pi}{n} j\right) &=& 
	\cos(\frac{2 \pi}{n} j) v_x(\theta) -  
	\sin(\frac{2 \pi}{n} j) v_y(\theta) ,\nonumber \\
	v_y\left(\theta + \frac{2 \pi}{n} j\right) &=& 
	\sin(\frac{2 \pi}{n} j) v_x(\theta) +  
	\cos(\frac{2 \pi}{n} j) v_y(\theta),\nonumber \\
	p_x\left(\theta + \frac{2 \pi}{n} j\right) &=& 
	p_x \cos(\frac{2 \pi}{n} j) -
	p_y \sin(\frac{2 \pi}{n} j), \nonumber\\
	p_y\left(\theta + \frac{2 \pi}{n} j\right) &=& 
	p_y \cos(\frac{2 \pi}{n} j) +
	p_x\sin(\frac{2 \pi}{n} j)
\end{eqnarray} 
which means in particular that
\begin{eqnarray}
	A\left(\theta + \frac{2 \pi}{n} j \right)
	= A(\theta ).
\end{eqnarray}
We can insert this in the measure to obtain 
\begin{eqnarray}
	\bra{v_x}\ket{v_y} &=& 
	\int_0^{2 \pi} d \theta \, 
	A(\theta;\mu) v_x(\theta) v_y(\theta),\nonumber \\
	&=& \int_0^{\frac{2 \pi}{n}} A(\theta;\mu) \sum_{j=0}^{n-1} \frac{v_x^2(\theta) - v_y^2(\theta)}{2}\sin(\frac{4 \pi}{n}j) + 
	v_x(\theta)v_y(\theta) \cos(\frac{4 \pi}{n}j),\nonumber\\
	&=& 0.
\end{eqnarray}
In the final line, as explained we use
\begin{eqnarray}
	\sum_{j=0}^{n-1} \exp[i \frac{4 \pi}{n}j] = 0 \quad(n\geq 3)
\end{eqnarray}
Similarly identity also holds for the orthogonality relation to  particle density modes.
In summary, for $C_{2m\geq 4}$ system,  the orthogonal conditions are 
\begin{eqnarray}
    \bra{c}\ket{c} &= &\bra{p_i}\ket{p_i}=1 \nonumber\\
    \bra{c}\ket{p_i}&=&\bra{c}\ket{v_i} = 0\nonumber \\
    \bra{p_i }\ket{p_k } &= &
    \bra{p_i }\ket{v_k }=
    \bra{v_i }\ket{p_k } =
    \bra{v_i }\ket{v_k }= 0\quad (j\neq k).
    \label{eq:otho}
\end{eqnarray}
The collision operators can therefore be constructed as 
\begin{eqnarray}
	C_{\mathrm{mc}}\ket{h} &=& -\gamma_{\mathrm{mc}} {\ket{h}} +
	\gamma_{\mathrm{mc}} \bra{c}\ket{h} \ket{c} + \sum_i 
	\gamma_{\mathrm{mc}} \bra{p_i}\ket{h} \ket{p_i},\nonumber \\
	C_{\mathrm{mr}}\ket{h} &= &-\gamma_{\mathrm{mr}} {\ket{h}} +
	\gamma_{\mathrm{mr}} \bra{c}\ket{h} \ket{c},
\end{eqnarray}
which obey 
\begin{eqnarray}
\bra{c}C_{\mathrm{full}}\ket{h} &=& 0 \quad \text{Particle number conservation}, \nonumber \\
\bra{p_i}C_{\mathrm{mc}}\ket{h} &=& 0 \quad \text{Momentum conservation}.
\end{eqnarray}
This establishes an explicit  construction of the collision terms in the Boltzmann equations which obeys the necessary conservation laws. 

\section{Boltzmann equation and numeical solution}
The Boltzmann equation is 
\begin{eqnarray}
    \bigg[\partial_t + \dot{\mathbf{p}}\cdot \nabla_\mathbf{p} + \dot{\mathbf{x}} \cdot \nabla_\mathbf{r}\bigg] 
	f(\mathbf{x},\mathbf{p},t) = C[f].
\end{eqnarray}
For a single band topologically trivial system, the stationary solution satisfies
\begin{eqnarray}
    \bigg[\mathbf{F}\cdot \nabla_\mathbf{p} + \mathbf{v}(\mathbf{p})\cdot \nabla_\mathbf{r}\bigg] 
	f(\mathbf{x},\mathbf{p}) = C[f],
\end{eqnarray}
where $\mathbf{v}(\mathbf{p}) = \partial_\mathbf{p} \epsilon(\mathbf{p})$ and in the presence of electric field, $\mathbf{F} = -e \mathbf{E}$. In our calculation, we fix the longitudinal electric field $E_x$ and view $E_y(y)$ as a response to the input field $E_x$. 
Taking the linearized ansatz in Eq.~\eqref{eq:linear} and only considering linear response to the external field $E_x$, the Boltzmann equation for $h(y,\theta)$ reduces to the form mentioned in Eq.~(3) in the main text,
\begin{eqnarray}
    \partial_y \ket{v_y h} - E_x \ket{v_x} - E_y(y) \ket{v_y} &=& C_{\mathrm{full}}\ket{h}.
\end{eqnarray}

In order to solve the above equation numerically, we write 
\begin{eqnarray}
     \partial_y \ket{v_\mu h} + \gamma \ket{h}  = \underbrace{E_x \ket{v_x} +  E_y(y) \ket{v_y} + \gamma \braket{c}{h}\ket{c}
     + \sum_{i=1,2} \gamma_{\mathrm{mc}} \braket{p_i}{h}\ket{p_i}}_{S[h]}
     \label{eq:BTE}
\end{eqnarray}
In the absence of magnetic field, the BTE thus becomes 
\begin{eqnarray}
	\bigg( v_y   \partial_y + \gamma\bigg)h(y,\theta) 
	= [E_x v_x(\theta) + E_y(y)v_y(\theta) + \gamma h_0(y) Y_0(\theta) + \gamma_{\mathrm{mc}} 
	\sum_{i=1}^2 h_i(y) Y_i(\theta)]
	\equiv S(y,\theta).
\end{eqnarray}
Using a variable change to the characteristics,
\begin{eqnarray}
	\partial_s y &=& v_y, \quad y(0) = \mathrm{sgn}(\mathbf{v}_\theta \cdot \hat{y})\frac{w}{2} \nonumber \\
	\partial_s \theta &=& 0 , \quad \theta(0) = \xi,
\end{eqnarray}
the BTE simplifies to
\begin{eqnarray}
	(\partial_s + \gamma) h(s,\xi) =  S(s,\xi)
\end{eqnarray}

By choosing a fully diffusive boundary condition,
\begin{eqnarray}
	h(0,\xi) = 0.
\end{eqnarray}
the formal solution of the BTE reads
\begin{eqnarray}
	h(s,\xi) = \int_{0}^{\infty} ds'\, \Theta(s-s') e^{-\gamma(s-s')} S[h(s',\xi)].
	\label{eq:zerofieldsol}
\end{eqnarray}
The solution Eq.\,\eqref{eq:zerofieldsol} can be expressed in the original variables $h(y,\theta)$ via the transformation
\begin{eqnarray}
	\xi &=& \theta, \nonumber \\
	s &=& \frac{y+ \mathrm{sgn}(\mathbf{v}_\theta\cdot \hat{y})\frac{w}{2}}{\mathbf{v}_\theta \cdot \hat{y}}.
\end{eqnarray}
However, since the function $S$ depends explicitly on $h$, additional constraints need to be fulfilled to select a physical solution. The most important constraint is of course charge conservation,
\begin{eqnarray}
    \partial_t \rho + \nabla \cdot \mathbf{J} = 0
    \label{eq:charge}
\end{eqnarray}
We furthermore enforce the constraint $J_y(\pm W/2) = 0$ to ensure that no current can leak outside of the channel. For stationary states and zero magnetic field, this combines with Eq.\eqref{eq:charge} to give a vanishing $J_y$ for all values of $y$. 

Numerically, the solution to the BTE is constructed by iterating \eqref{eq:zerofieldsol} and imposing a projection of current $J_y$ to eliminate the transverse current as follows,
\begin{eqnarray}
    \ket{h^{\mathrm{New}}} = (\mathbb{1} - \ket{v_y} \bra{v_y})\ket{h^{\mathrm{Old}}}.
\end{eqnarray}
A good starting point is to assume $S(y,\theta) = E_x v_x(\theta)$. 
Then 
\begin{eqnarray}
	h^0(y,\theta) = 
	\frac{E_x (\mathbf{v}_\theta \cdot \hat{x}) }{\gamma}\bigg[1 - \exp\left(-\gamma\frac{2\,\mathrm{sgn}(\mathbf{v}_\theta\cdot \hat{y})y+ w}{2|\mathbf{v}_\theta \cdot \hat{y}|}\right)\bigg].
 \label{eq:guess}
\end{eqnarray}
The numerical flow chart is summarized in Fig.~\ref{fig:flowchart}.
\begin{figure}
	\centering
\begin{tikzpicture}[thick,scale=0.55, every node/.style={scale=0.8}]
	\node[draw, align=center,rounded corners]        (input)  at (0,0) {Input parameters \\
	$\gamma_{\mathrm{mr}},\gamma_{\mathrm{mc}},B,E_x $ \\
	Input initial guess/last epoch result\\
	$h_{\text{last}}(y,\tau) / S(y,\tau)$};

	\node[draw,below=20pt of input,align=center]
	(interp) {
		Interpolation of $h(y,\tau)$  \\
	$h(y,\tau) \Rightarrow h(y(s,\xi),\tau(s,\xi))$};

	\node[draw,below=60pt of input,align=center]
	(C0) {
		Inversion of $C_0$ and projection \\
	$\hat{h}(s,\xi) = \hat{P}C_0^{-1}[S(s,\xi)]$};

	\node[draw, diamond,aspect=2, below=100pt of input]           
	(conv check)  {$||\hat{h}(y,\theta) - h(y,\theta)|| < \epsilon$?};

	\node[draw, right=20pt of conv check] (update) {$h(y,\theta) = (1-\eta)h(y,\theta) + \eta \hat{h}
	(y,\tau)$};

	\node[draw, above=20pt of update, align=center] (source) {
	Calculation of source terms from $h(y,\theta)$ \\	
	$h(y,\theta) \Rightarrow S(y,\theta) $};

	\node[draw,below=30pt of conv check, align=center](output){
		Output result and start new loop \\
		$h_{\mathrm{out}} = h(y,\theta)$
	}; 

	
	
	\draw[-{Latex[length=1.5mm]}] (input)  -- (interp);
	\draw[-{Latex[length=1.5mm]}] (interp)  -- node[left]{} (C0);
	\draw[-{Latex[length=1.5mm]}] (C0)  -- node[left]{} (conv check);
	\draw[-{Latex[length=1.5mm]}] (conv check)  -- node[above]{No}  (update);
	\draw (input) -- coordinate (mid1) (interp);
	\draw[-{Latex[length=1.5mm]}] (update)  -- (source);
	\draw[-{Latex[length=1.5mm]}] (source) |- (mid1);
	\draw[-{Latex[length=1.5mm]}] (conv check)  -- node[right]{Yes}  (output);
	
  \end{tikzpicture}
  \caption{Flow chart of the numerical routine used to solve the BTE.}
  \label{fig:flowchart}
\end{figure}
As a benchmark, we compare our calculation result to Ref.~\cite{deJong1995} in Fig.~\ref{S4}. The agreement is excellent in the Gurzhi region, but we note that the solver was not optimized for the deep hydrodynamic limit, where it converges very slowly. 
\begin{figure}
    \centering
    \includegraphics[width =  \textwidth]{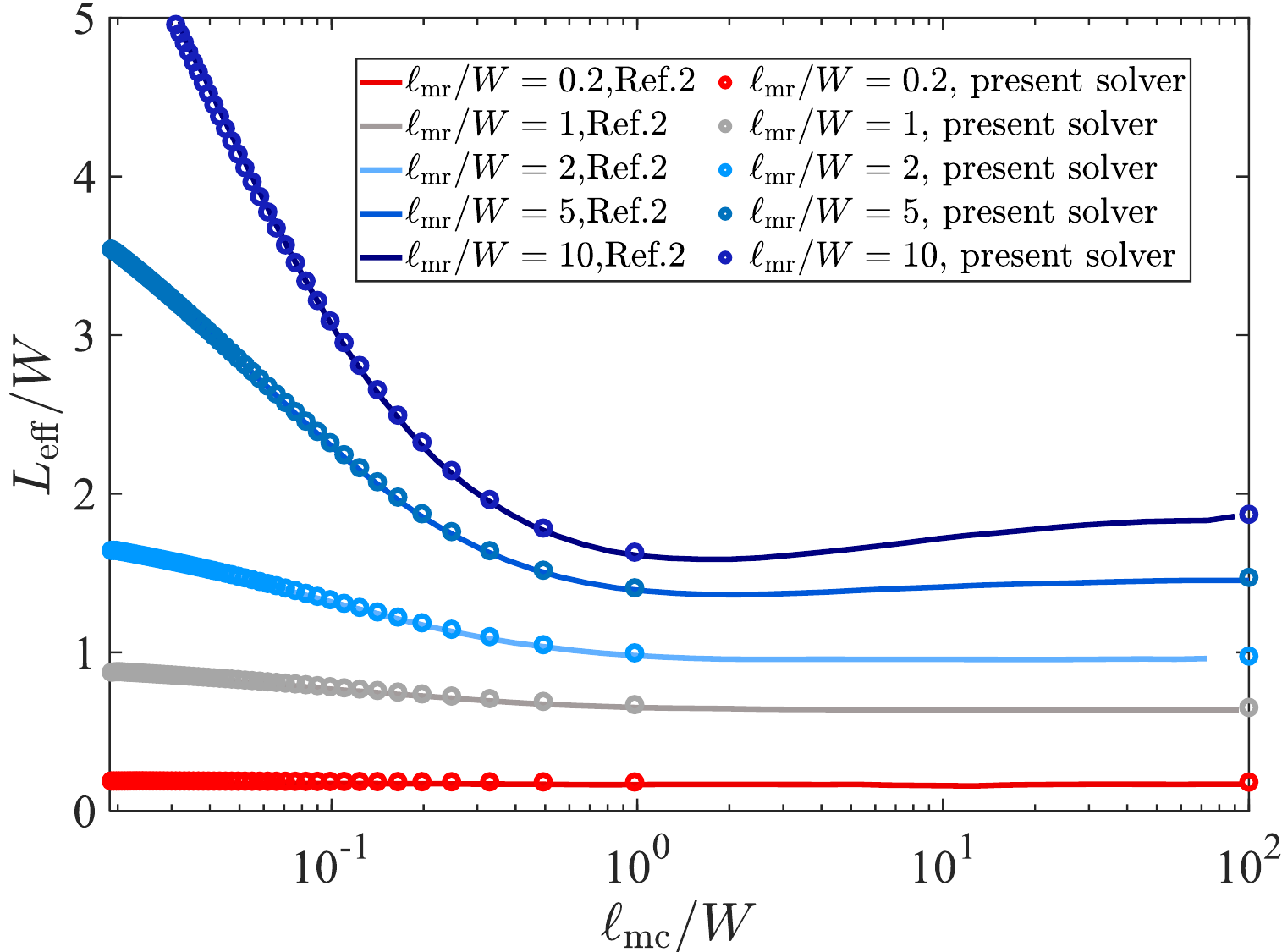}
    \caption{Comparision between our BTE solver to the previous results in Ref.~\cite{deJong1995}.}
    \label{S4}
\end{figure}

\section{Transverse voltage decomposition}
In order to obtain transverse voltage, we can either act with the bra $\bra{v_y}$ or alternatively with $\bra{p_y}$ from the left in Eq.~\eqref{eq:BTE}. Utilizing Eq.~\eqref{eq:otho} one can then immediately solve for $E_y$. 

By contracting with $\bra{p_y}$, we obtain 
\begin{eqnarray}
    E_y &=& \frac{\partial_y \expval{p_y v_y} + \gamma_{\mathrm{mr}} \expval{p_y}}{\bra{p_y}\ket{v_y}} \\
    &\equiv& \frac{\partial_y \Pi_{y y} + \gamma_{\mathrm{mr}} \,p_y(y)}{\bra{p_y}\ket{v_y}},
    \label{eq:decomhE}
\end{eqnarray}
where $\Pi$ is the stress tensor of a quantum fluid~\cite{Bradlyn2012,massignan2005viscous} and $p_y(y)$ is the momentum density. 
Integrating over $y$, we yield the transverse voltage
\begin{eqnarray}
    \Phi_{y} = \frac{1}{{\bra{p_y}\ket{v_y}}} \bigg( \Pi_{yy}\bigg|_{-W/2}^{W/2}
    +  \gamma_{\mathrm{mr}} P_y \bigg).
    \label{eq:decomh}
\end{eqnarray}
This result clearly shows that in the hydrodynamic limit only the internal stress contributes to the transverse voltage, i.~e. ($\gamma_{\mathrm{mr}}/\gamma_{\mathrm{mc} }\rightarrow 0$), 
\begin{eqnarray}
    \Phi_y^{\mathrm{hydro}} = \frac{1}{{\bra{p_y}\ket{v_y}}}  
    \Pi_{yy}\bigg|_{-W/2}^{W/2}.
\end{eqnarray}

Interestingly, the alternative way of decomposing $\Phi_y$ gives rise to the ballistic limit: By contracting with $\bra{v_y}$, one obtains
\begin{eqnarray}
    E_y &=& \frac{\partial_y \expval{v_y v_y}-\gamma_{\mathrm{mc} }\expval{p_y}}{\bra{v_y}\ket{v_y}} \\
    &\equiv& \frac{ \partial_y {\Sigma}_{y y} -\gamma_{\mathrm{mc} }\, p_y(y) }{\bra{v_y}\ket{v_y}}, 
\end{eqnarray}
where $\Sigma_yy=\expval{v_y v_y}$ is the velocity-velocity correlator. 
Integrating over $y$, this becomes
\begin{eqnarray}
    \Phi_{y} = \frac{1}{{\bra{v_y}\ket{v_y}}} \bigg( \Sigma_{yy}\bigg|_{-W/2}^{W/2}
    -  \gamma_{\mathrm{mc}} P_y \bigg)
    \label{eq:decomb}
\end{eqnarray}
As promised, in the ballistic limit ($\gamma_{\mathrm{mc}}=0$),  the transverse voltage is given by
\begin{eqnarray}
      \Phi_y^{\mathrm{ball}} = \frac{1}{{\bra{v_y}\ket{v_y}}}  
    \Sigma_{yy}\bigg|_{-W/2}^{W/2}.
    \label{eq:ball}
\end{eqnarray}

In the main text, we opted for the decomposition according to Eq.~\eqref{eq:decomh} rather than Eq.~\eqref{eq:decomb}. The reason is straightforward, compared to the correlator $\Sigma$, whose properties are not well documented, the stress tensor $\Pi$ has a clear physical interpretation.
In particular, $\Pi$ allows us to formulate a continuity equation for momentum $p_y(y)$, which reads
\begin{eqnarray}
    \frac{\partial p_y}{\partial t} 
    = - \frac{\partial \Pi_{ij}}{\partial x_j} -
    \gamma_{\mathrm{mr}}{p_y} + eE_y
    \label{eq:macro}
\end{eqnarray}
In the steady state, the temporal derivative in \eqref{eq:macro} vanishes, so that
\begin{eqnarray}
    e E_y = \gamma_{\mathrm{mr}} p_y(y) + 
    \partial_y \Pi_{yy}(y)
\end{eqnarray}
which leads us immediately to \eqref{eq:decomhE} once the proper normalization is imposed. 

\section{Symmetric and anti-symmetric component of transverse voltage}

For Fermi surface with $C_{2m}$ symmetry, the irreducible misalignment angle domain is $\varphi \in [0,\pi/2 m]$ since $\Phi_y(\varphi + \pi/2m)
 = -\Phi_y(\varphi + \pi/2m)$. The anisotropic nature of Fermi surface results in the $\Phi_y$ v.s. $\varphi$ profile asymmetric, i.e. the maximum value of $\Phi_y$ does not coincide with $\varphi = \pi/4m$(cf. main text Fig.2(c)). We then further look into the the symmetric and anti-symmetric component of $\Phi_y$ w.r.t. $\varphi$ in $ [0,\pi/2 m]$. They are defined as follows:
 \begin{eqnarray}
     \Phi_y^a(\varphi) &=&  \frac{1}{2}[\Phi_y(\varphi) - \Phi_y(\pi/2 m-\varphi)]\nonumber \\
     \Phi_y^s(\varphi) &=& \frac{1}{2}[\Phi_y(\varphi) + \Phi_y(\pi/2 m-\varphi)]
 \end{eqnarray}

In Fig.~\ref{S6}(a), we showed that for $\mathrm{PdCoO_2}$, the anti-symmetric component is more susceptible to MC scattering and changes sign with smaller $\gamma_{\text{mc}}v_F/W\approx 0.2$. In  Fig.~\ref{S6}(b) and (c), we clearly see that for $\gamma_{\text{mc}}v_F/W<0.2$, the symmetric part is dominant and the whole transverse voltage does not change sign.  

\begin{figure}[H]
    \centering
	\includegraphics[width=\linewidth]{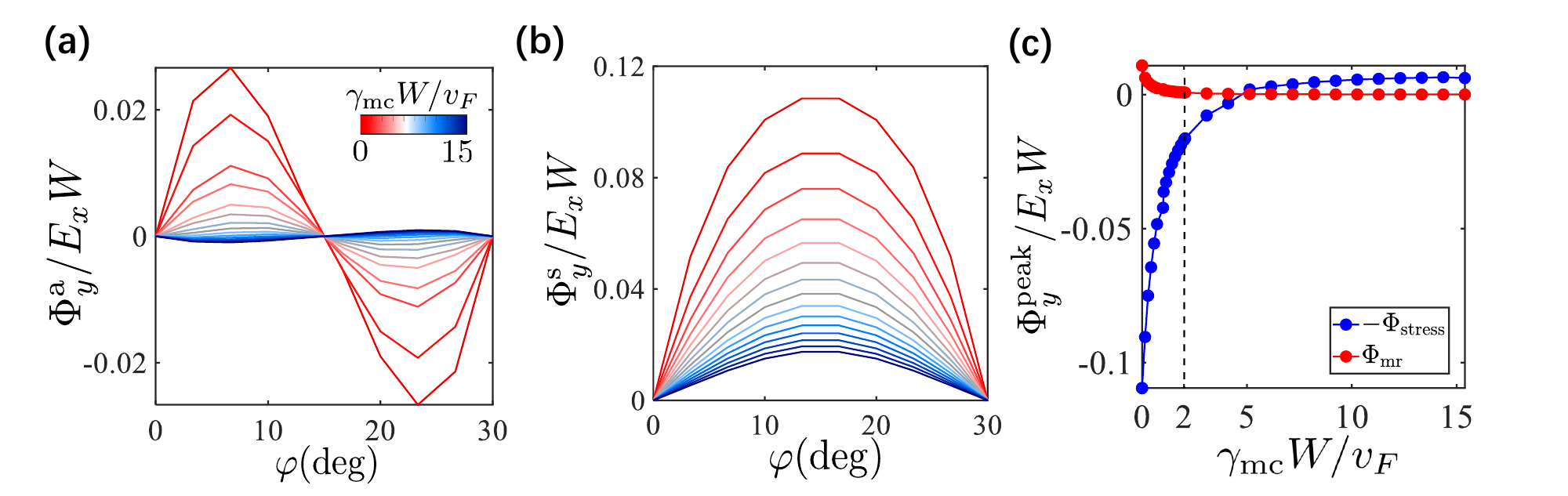}
	\caption{Transverse voltage plots for $\mathrm{PdCoO_2}$ with same parameters as in Fig.4(b) in main text. (a)Anti-symmetric component of transverse voltage. (b)Symmetric component of transverse voltage. (c)Full transverse voltage decomposition. Black dash line is an eye guide to $\gamma_{\text{mc}}v_F/W = 0.2$. }
    \label{S6}
\end{figure}

This observation helps to identify a weaker hydrodynamic signature in real materials. The sign change of the anti-symmetric component of the transverse voltage could serve as an effective detector for broader experimental materials due to its earlier onset with the MC scattering rate.

\section{Extreme ballistic limit}
\begin{figure}[H]
    \centering
	\includegraphics[width=0.35 \linewidth]{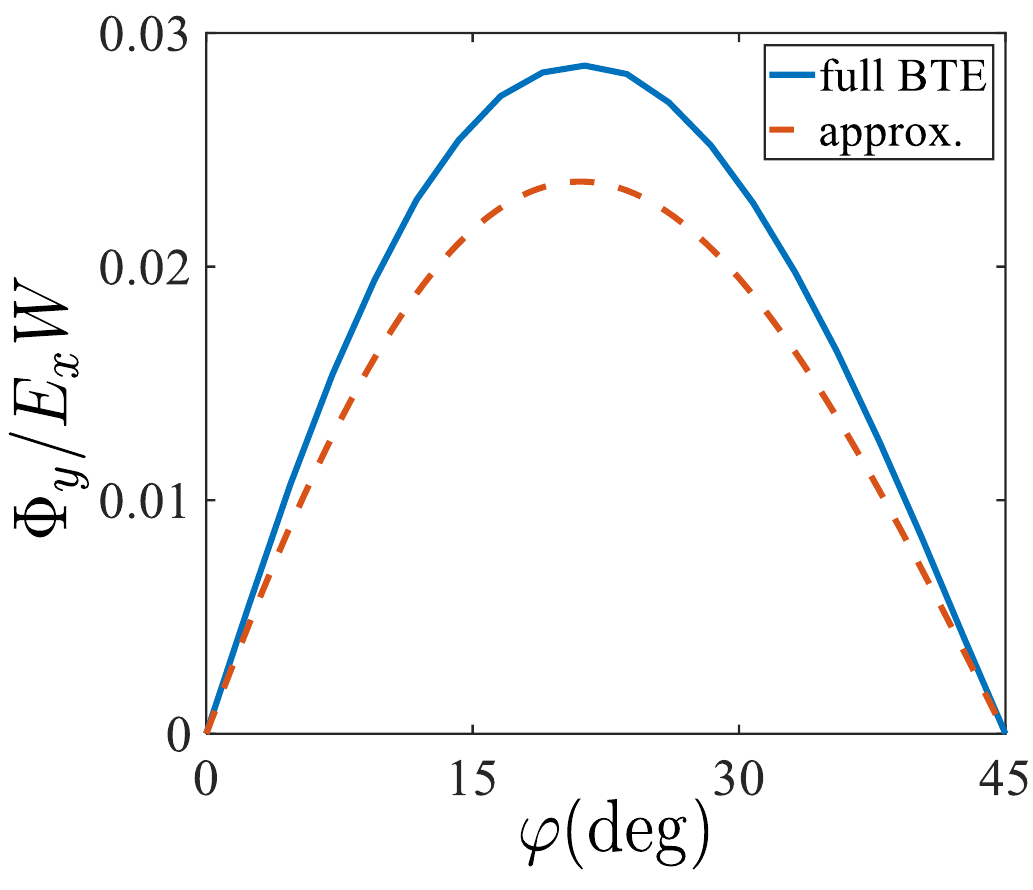}
	\caption{Comparison between a full solution and approximated analytic result,$\alpha=0.4$.}
    \label{S1}
\end{figure}

\begin{figure}[H]
    \centering
	\includegraphics[width=0.35 \linewidth]{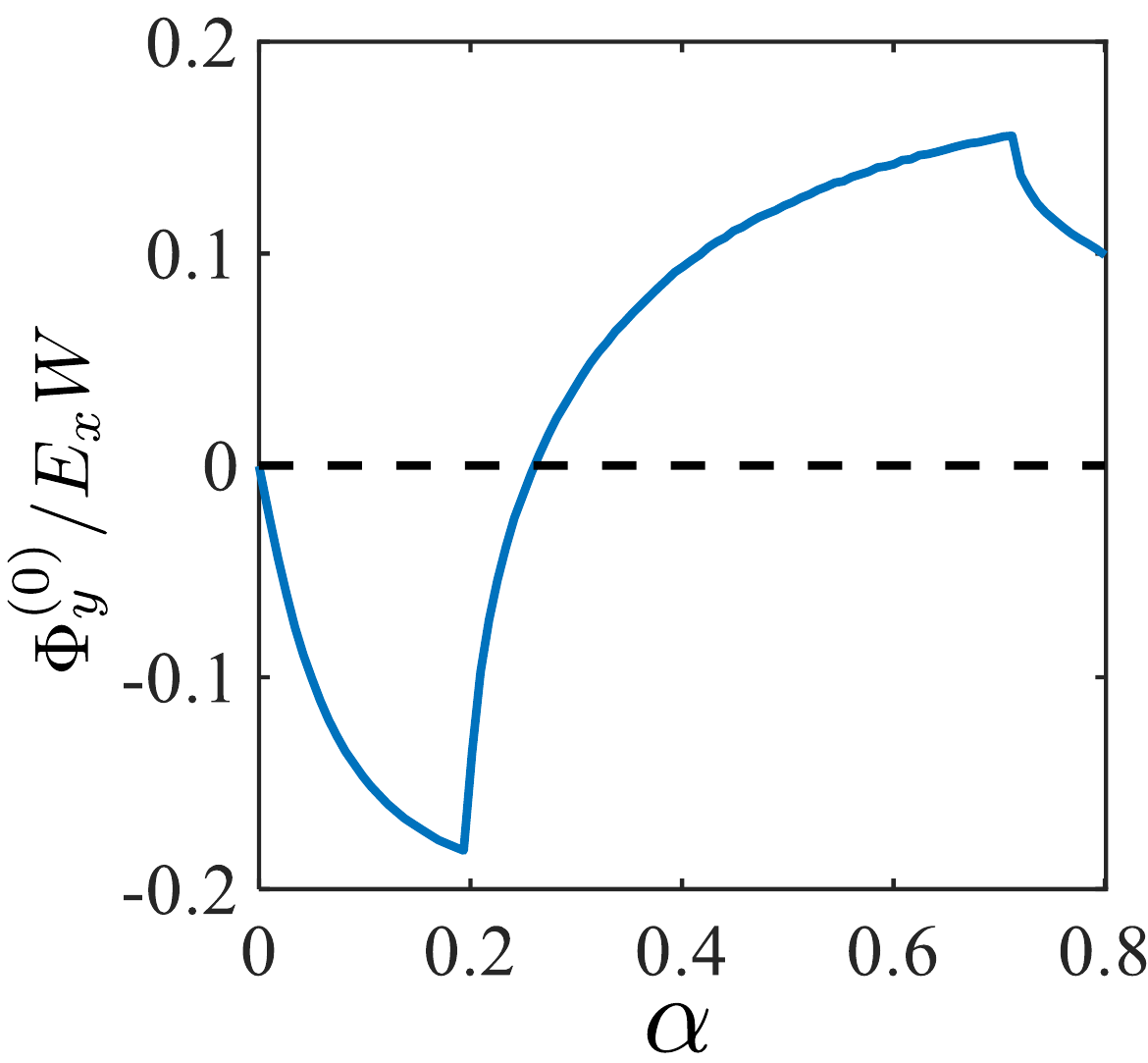}
	\caption{Transverse voltage according to Eq.~\eqref{eq:simp} based on the quantity $\mathrm{sign}(v_y)v_x$. Parameters are the same as in main text Fig. 2(b).}
    \label{S2}
\end{figure}

Here we show the results of a pure ballistic case and further explain Fig.2(b) in the main text. For ballistic case, $\gamma = \gamma_{\mathrm{mr}}$, and the source term $S$ is constituted only from electric field and particle number modes. As it turns out, we find that already for not too small $\gamma_{\mathrm{mr}}$, solution ~\eqref{eq:guess} is a good approximation to the full solution of distribution function. 
According to Eq.~\eqref{eq:ball}, the transverse voltage can be written explicitly as
\begin{eqnarray}
    \Phi_y^{(0)} =\frac{2 E_x}{\gamma}  \frac{
\int d \theta A(\theta) v_y^2(\theta) v_x(\theta) e^{-\gamma w/2 |v_y|} \sinh\left( \frac{\gamma W}{2 v_y}\right)}{\int d \theta A(\theta) v_y^2(\theta)}.
\label{eq:phiy0}
\end{eqnarray}
As shown in Fig.~\ref{S1}, this approximation and the numerical solution show the same qualitative behavior, meaning that Eq.~\eqref{eq:phiy0} can successfully capture \textbf{the sign} of the voltage. 
For very small $\gamma$, Eq.~\eqref{eq:phiy0} reduces to 
\begin{eqnarray}
    \Phi_y^{(0)} &=&- \frac{\gamma}{2}E_x W^2 \frac{\int d\theta A(\theta) v_x \mathrm{sign}(v_y)}{\int d\theta A(\theta) v_y^2(\theta)},
    \label{eq:simp}
\end{eqnarray}
which allows us to read off what decides the sign of transverse voltage in the extreme ballistic limit, which is the average $\expval{\mathrm{sign}(v_y)v_x}$. The physical content in this average is as follows: For a given electric field $E_x$, the response to it is trivially proportional to $v_x$, while $\mathrm{sign}(v_y)$ counts whether the electrons move to the top/bottom in the transverse direction of the channel. The resulting voltage plot as a function of $\alpha$ is shown in Fig.~\ref{S2}, where the sign change is at $\alpha_c \approx 0.3$, very similar to the sign change shown in Fig.2(b) of the main text, which was obtained from the full calculation. 

\section{Transverse voltage and longitudinal conductance for {{$D_4$}} model}

\begin{figure}[ht]
	\includegraphics[width= \linewidth]{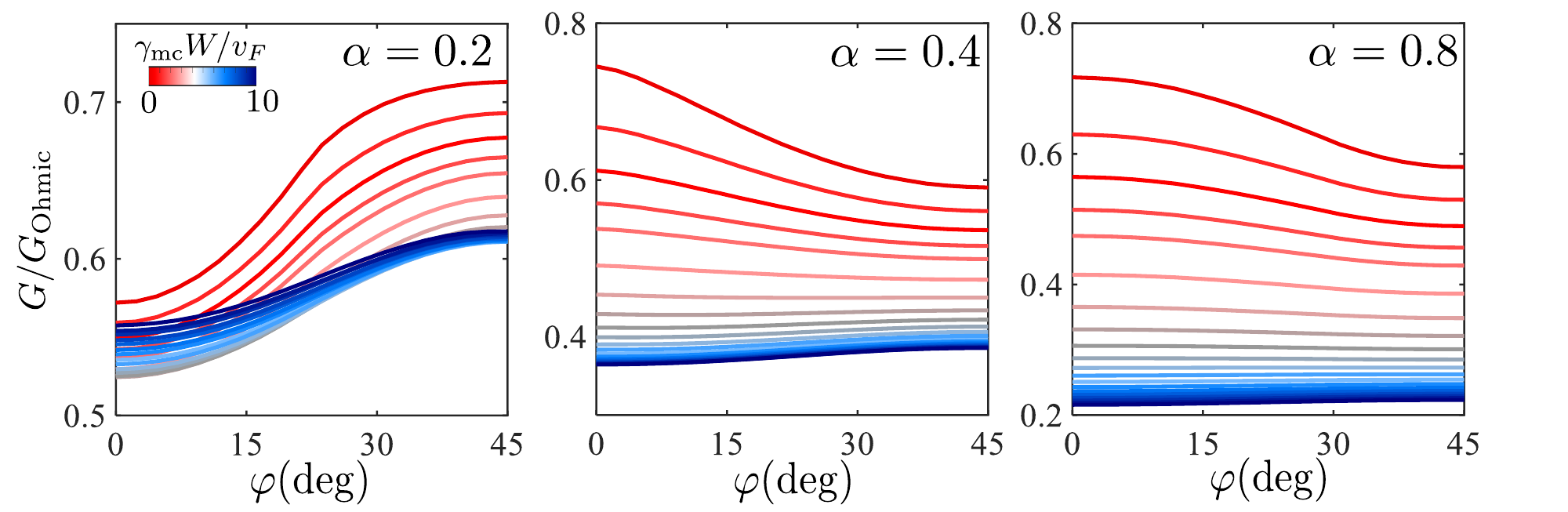}
	\caption{Conductance $G$ as a function of different misalignment angles $\varphi$ and MC scattering rate $\gamma_{\mathrm{mc}}$ for different $\alpha$. Here, all $\gamma_{\mathrm{mr}}W/v_F=1$.}
    \label{S3}
\end{figure}

Here we show the longitudinal conductance $G$ for our {{$D_4$}} model with different Fermi surface anisotropy $\alpha$ in Fig.~\ref{S3}. The conductance $G$ is defined as 
\begin{eqnarray}
   G = \int d y J_x(y)/E_x. 
\end{eqnarray}

Firstly, the conductance is varied with changing misalignment angles, showing its extreme values at high symmetry positions ($\varphi=0^\circ,45^\circ$). We observed quite different behavior of the conductance with varying Fermi surface anisotropy. Looking at $\alpha = 0.4,0.8$, from the ballistic to the hydrodynamic limit, the conductance at a fixed misalignment angle decreases consistently. 
On the other hand, $\alpha=0.2$ leads to crossings between different lines, meaning different misaligned configuration lead to different relationships with respect to $\gamma_{\mathrm{mc}}$.  It is evident that the anisotropy of the Fermi surface leads to different scaling behaviors (at least scaling with $\ell_{\mathrm{mc}}$) for different Fermi surfaces. 
This finding further supports the conclusions drawn in the main text that anistropic Fermi surfaces can lead to qualitatively different flow behavior, and it calls into question how generally applicable some of the results for an isotropic model are when compared to real materials.

{We also find non-monotonic behavior for the transverse voltage as a function of $\gamma_{\mathrm{mc}}$ for certain $D_4$ Fermi surfaces, aligning with the Gurzhi effect as a crossover from ballistic to hydrodynamic regime. As shown in Fig.~\ref{S6}, the transverse voltage of $\alpha=0.31$ displays a non-monotonic behavior while the $\alpha = 0.38$ shows a monotonic decrease with $\gamma_{\mathrm{mc}}$. This is related to the transition from Knudsen to Poiseuille regime~\cite{deJong1995,Heilmann2024}. }

\begin{figure}[ht]
	\includegraphics[width= 0.4 \linewidth]{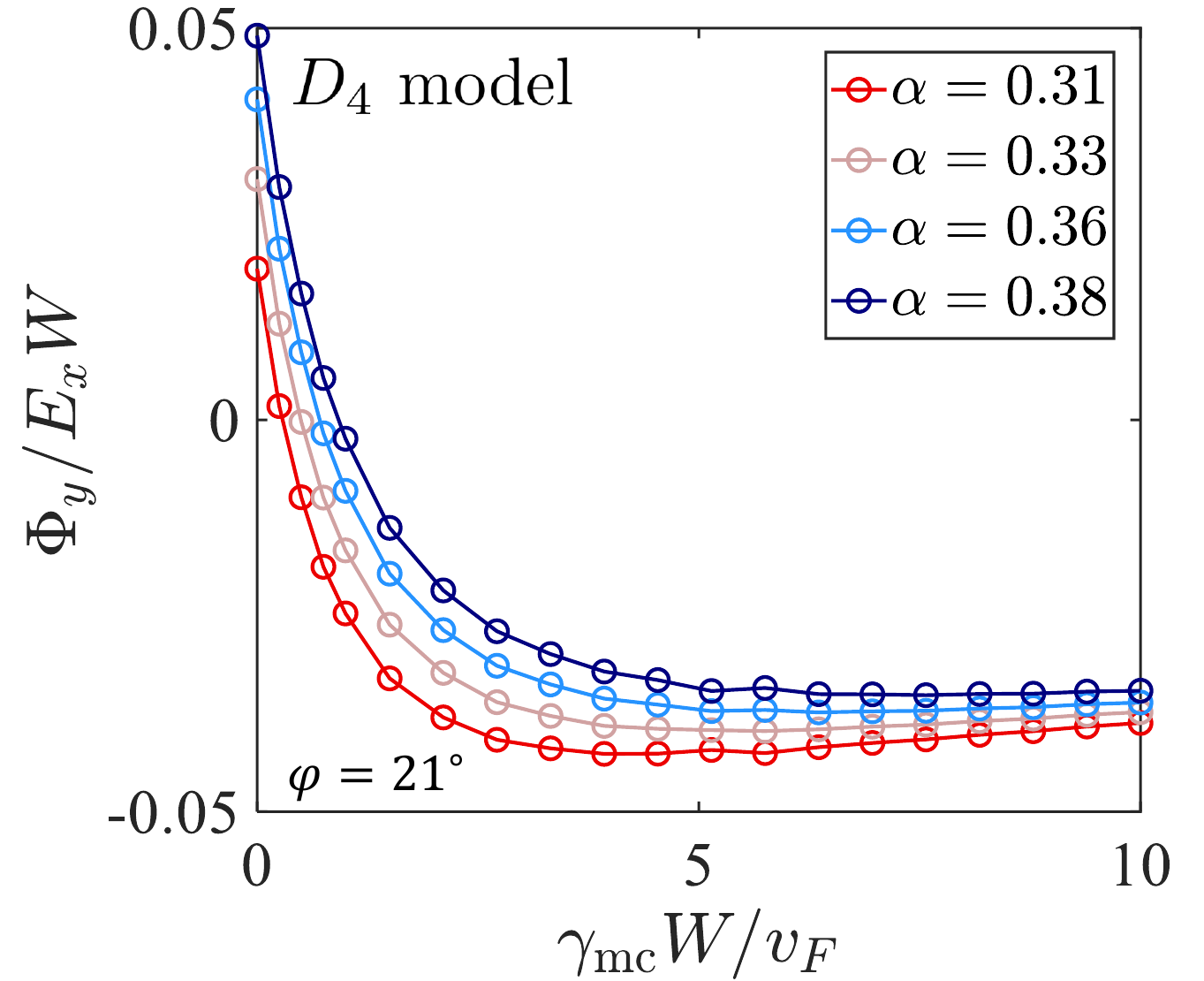}
	\caption{Transverse voltage at misalignment angle $\varphi = 21^\circ$ for different Fermi surfaces.}
    \label{S6}
\end{figure}

\section{{Influence of defects}}
\begin{figure}[ht]
	\includegraphics[width=\linewidth]{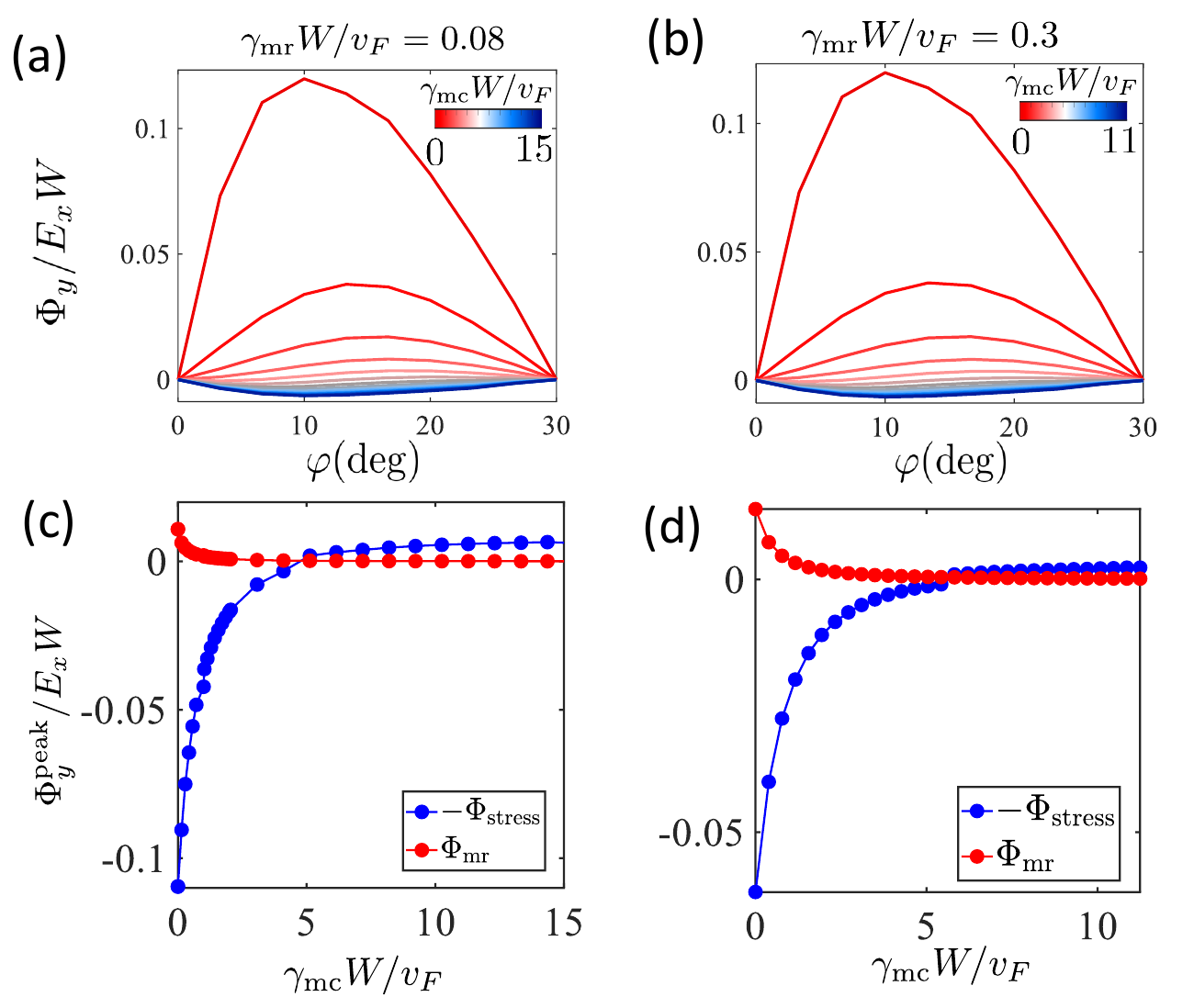}
	\caption{{Transverse voltage of $\mathrm{PdCoO_2}$ model with different $\gamma_\mathrm{mr}$. (a)-(b): Total transverse voltage as a function of misalignment angle and momentum-conserving scattering rate. (c)-(d): Transverse voltage decomposition of the two cases. The parameters of (a) and (c) are the same as in the main text. }}
    \label{S7}
\end{figure}

{
Here we discuss the sensitivity of our transverse voltage sign change phenomena against different strength of defects by comparing the transverse voltage under different $\gamma_{\mathrm{mr}}$ in the $\mathrm{PdCoO_2}$ model. 

As shown in Fig.~\ref{S7}, the transverse voltage sign change also happens at larger impurity scattering strength, but this requires larger $\gamma_{\mathrm{mc}}$ to enable the crossover. It is also worth noting that the signal is smaller in the larger defect case. }

\section{Continuous and tight-binding models for realistic systems}
The models used in main text Fig.4  of realistic systems are presented here. 

\paragraph{The continuous {$D_4$} model} 
is a real band dispersion for the geometrical $C_4$ model discussed in Fig.(2) of main text.  The energy dispersion is $\epsilon(\mathbf{k})={k^2}/{2 m} - 2 t ({k_x^4 + k_y^4 - 6 k_x^2 k_y^2})/{k^4} - \mu$ with parameters $m = 0.5,t=1,\mu =1$. 

\paragraph{The Square model} is a simple tight-binding model at nearly half-filled. The energy dispersion is $\epsilon(\mathbf{k})= -2t(\cos k_x + \cos k_y) - \mu$ with parameters  $t=1,\mu =-0.1$.

\paragraph{The $\mathrm{PdCoO_2}$ model} describes the in-plane dispersion of $ \mathrm{PdCoO_2}$, ignoring the small corrugation along $k_z$ direction~\cite{takatsu2013extremely}. 
The energy dispersion is   
\begin{align}
\epsilon(\mathbf{k})=  -2 t_1\{\cos (\boldsymbol{k} \cdot \boldsymbol{a})+\cos (\boldsymbol{k} \cdot \boldsymbol{b})+\cos [ \boldsymbol{k} \cdot(\boldsymbol{a}+\boldsymbol{b})]\} 
-2 t_2\left\{\cos ^2(\boldsymbol{k} \cdot \boldsymbol{a})+\cos ^2(\boldsymbol{k} \cdot \boldsymbol{b})\right. 
 \left.+\cos ^2[ \boldsymbol{k} \cdot(\boldsymbol{a}+\boldsymbol{b})]\right\},
\end{align}
with parameters  $\mathbf{a} = \hat{x} , \mathbf{b} = \frac{1}{2} (\hat{x} + \sqrt{3} \hat{y}), t_1=1, t_2 =0.14$.

\paragraph{The $\mathrm{Bi2212}$ model} describes the in-plane dispersion of overdoped cuprate $\mathrm{Bi_2Sr_2CaCu_2O_8}$ , ignoring the small corrugation along $k_z$ direction~\cite{markiewicz2005one}. For convenience, define  $c_i(\alpha a) = \cos(\alpha k_i a)$.  The 
energy dispersion is   $\epsilon(\mathbf{k})=$   $   -2 t_1\left[c_x(a)+c_y(a)\right]-4 t_2 c_x(a) c_y(a)-2 t_3\left[c_x(2 a)\right.  
\left.+c_y(2 a)\right]-2 t_4\left[c_x(2 a) c_y(a)+c_y(2 a) c_x(a)\right]
    $ with parameters $a=1,t_1 = 1,t_2 = -0.135, t_3 = 0.061,t_4 = -0.017$.

\bibliography{literature}